\documentclass[prb,twocolumn]{revtex4}

\usepackage{epsf}
\usepackage{epsfig}
\usepackage{subfigure}
\usepackage{textcomp}
\usepackage{dcolumn}
\usepackage{bm}

\begin{document}


\title{Spin-orbit interactions and spin-currents from an exact-exchange 
Kohn-Sham method}

\author{Stefan Rohra$^1$, Eberhard Engel$^2$ and Andreas G\"orling$^1$}
\affiliation{$^1$Lehrstuhl f\"ur Theoretische Chemie,
Universit\"at Erlangen-N\"urnberg, Egerlandstr. 3, D-91058 Erlangen, Germany}
\affiliation{$^2$Institut f\"ur Theoretische Physik,
J.W.Goethe-Universit\"at Frankfurt, 
Max-von-Laue-Str.\ 1, D-60438 Frankfurt/Main, Germany}

\date{\today}

\begin{abstract}
An exact-exchange spin-current Kohn-Sham method to treat non-collinear spin,
magnetic effects, currents, spin-currents, and spin-orbit interactions
self-consistently on equal footing is introduced. Spin-orbit interactions are
shown to induce spin-currents. Results for silicon and germanium are
presented.
\end{abstract}


\maketitle

\def\angst{\,\text{\AA}}

\newcommand{\elmat}[3]{\langle {#1} | {#2} |{#3} \rangle}

\def\RR{({\bf r})}
\def\RB{{\bf r}}       
\def\RP{({\bf r}')}
\def\RPP{({\bf r}'')}
\def\phiG#1{\phi^{\Gamma\!,\gamma}_{#1}}
\def\phiGT#1{{\phi^{\Gamma\!,\gamma}_{#1}}^T\!\!}
\def\phiU#1{\phi^{\Upsilon\!,\upsilon}_{#1}}
\def\phiGP#1{\phi^{\Gamma'\!,\gamma'}_{#1}}
\def\phiGPT#1{{\phi^{\Gamma'\!,\gamma'}_{#1}}^T\!\!}
\def\phiGM#1{\phi^{\Gamma_M,\gamma}_{#1}}
\def\phiGMT#1{{\phi^{\Gamma_M,\gamma}_{#1}}^T\!\!}
\def\phiGm#1{\phi^{\Gamma\!,\mu}_{#1}}
\def\phiGmT#1{{\phi^{\Gamma\!,\mu}_{#1}}^T\!\!}
\def\nG#1{n^{\Gamma}_{#1}}
\def\nGP#1{n^{\Gamma'}_{#1}}
\def\nGM#1{n^{\Gamma_M}_{#1}}
\def\phiL#1{\phi^{\Lambda,\lambda}_{#1}}
\def\phiLT#1{{\phi^{\Lambda,\lambda}_{#1}}^T\!\!}
\def\nL#1{n^{\Lambda}_{#1}}
\def\gG{g^{\Gamma}}
\def\gGP{g^{\Gamma'}}
\def\gGM{g^{\Gamma_M}}
\def\gL{g^{\Lambda}}
\def\eG#1{\varepsilon^{\Gamma}_{#1}}

\def\coulP{|{\bf r}'-{\bf r}''|}
%

%
%


The electronic structures of crystalline materials containing 
heavier elements 
are strongly affected by spin-orbit effects. 
Even in the band structure of germanium, a third row element, 
spin-orbit splittings of 
about 0.25 eV occur. Thus for an accurate description of electronic structures
of most interesting materials a treatment of spin-orbit effects is highly 
desirable. If not just simple bulk semiconductors but also other 
materials, e.g., oxides or metals or one- and two-dimensional periodic systems
like molecular wires or surfaces, shall be considered then in addition to
spin-orbit
effects also spin-polarization, including non-collinear spin-polarization,
needs to be treated.
If magnetic properties shall be studied, an inclusion
of currents is necessary. In the emerging field of spintronics besides density
currents also spin currents may be of importance. Materials with properties of
technical interest, e.g. for new semiconductors or magnetooptical storage
devices or for spintronics, often are characterized by an interplay of
spin-orbit interactions, non-collinear spin-polarization, orbital currents, and
spin-currents. Therefore it seems highly desirable to develop an electronic
structure method that treats all of the mentioned effects in a unified way on
equal footing. This work outlines such a method 
within the framework of density-functional theory (DFT) and, as a central 
result, shows how spin-orbit interactions generate spin-currents 
and how the latter can be treated in a self-consistent manner.
As an illustration of the new method we calculate
spin-orbit effects in the bandstructures of silicon and germanium with special
emphasis on spin-currents induced by spin-orbit effects. 

DFT is applied almost exclusively within the Kohn-Sham (KS) formalism 
\cite{yang89}. 
In recent years KS methods that treat the exchange energy 
as well as the local multiplicative
KS exchange potential exactly were developed 
\cite{kotani,stadele97,gorling05,kurth}. 
Such exact-exchange (EXX) Kohn-Sham (KS) methods proved to describe 
the electronic structure of periodic systems, in particular 
bandstructures and band gaps of semiconductors, distinctively better 
than conventional KS methods based on the local density approximation (LDA) 
or generalized gradient approximations (GGAs)\cite{stadele97}. 
%
%
 
EXX KS methods not only yield improved bandstructures. Furthermore, they
represent a systematic improvement over LDA and GGA methods in the sense that
the largest fraction of the exchange-correlation functionals for energy 
and potential, i.e., the exchange part, no longer needs to be approximated but
is treated exactly. This is crucial for the treatment of
magnetic effects within DFT. Already about 20 years ago Vignale and Rasolt 
introduced current-spin-density-functional theory \cite{vignale87,vignale88} 
which, in principle,
enables a treatment of both spin and magnetic effects. 
In practice, however, current-spin-density-functional theory
could not be applied to realistic systems because 
reliable and tractable approximate current-density functionals were not
available. The EXX framework offers the opportunity to avoid at least exchange
approximations and thus opens the route to  current density functional
methods \cite{helbig01,rohra06}. 
Indeed, we recently presented an EXX spin-current
density-functional method that is based on a spin-current
density-functional theory (SCDFT) \cite{rohra06}. SCDFT generalizes the 
current spin-density-functional theory of Vignale and Rasolt
by taking into account spin-currents in addition to 
the quantities considered in
current spin-density-functional theory, i.e., the electron density, 
the x-, y-, z-components of the spin-density, 
and the x-, y-, z-components of the paramagnetic current density. The
spin-currents that are additionally considered in SCDFT 
are the x-, y-, z-components of the paramagnetic 
currents of the x-, y-, z-components 
of the spin-density, i.e., paramagnetic currents of the magnetization.

The EXX SCDFT method enables the treatment of non-collinear spins,
currents of the electron density and, going beyond 
current-spin-density-functional theory, spin-currents. By introducing in this
work spin-orbit and scalar relativistic effects in the EXX SCDFT method
the desired comprehensive description of electronic and magnetic properties 
of materials including relativistic
effects, in particular spin-orbit effects, becomes possible. 

Our starting point, the EXX SCDFT method of Ref.\ [\onlinecite{rohra06}], 
employs basis sets of
plane waves and thus requires to treat core electrons via
pseudopotentials. Relativistic effects including spin-orbit splittings
are generated almost exclusively in the region close to nuclei. In this region
orbitals and subsequently the electronic structure is determined by the
pseudopotential. In order to introduce relativistic effects, we therefore
introduce spin-orbit-resolved normconserving EXX 
pseudopotentials $v_{\ell j}^{PP}$ 
with $\ell$ denoting the spatial orbital momentum and $j=\ell\pm 1/2$
($j=1/2$ for $\ell=0$) denoting the 
total, i.e., spatial plus spin, angular momentum. 
Instead of nonrelativistic $\ell$-dependent EXX
$s$-,$p$-,$d$-pseudopotentials we thus construct $\ell j$-dependent EXX
$s_{1/2}$-,$p_{1/2}$-,$p_{3/2}$-,$d_{3/2}$,$d_{5/2}$-pseudopotentials that
take into account all relativistic effects and
employ them as separable pseudopotentials 
\begin{eqnarray}
\label{pseudopotential}
v_{lj}^{sep} \!\!\!\!&(&\!\!\!\!{\bf r,r'}) =
\sum_{m=-j}^{j} \Big[ \int dr r^2 \phi_{lj}^\dagger(r) v_{lj}^{PP}(r)
\phi_{lj}(r) \Big]^{-1}
\! \times \nonumber \\
&\times& \!\!
v_{lj}^{PP}(r) \phi_{lj}(r) \Omega_{ljm} ({\hat{\bf r}})
v_{lj}^{PP}(r') \phi_{lj}(r') \Omega_{ljm}^\dagger ({\hat{\bf r}}').
\end{eqnarray}
In Eq.(\ref{pseudopotential}) the $\Omega_{ljm}$
denote central field spinors, i.e., two-component angular momentum 
eigenfunctions with total angular momentum $j$ and corresponding magnetic
quantum number $m$ that result from angular momentum coupling of spin 
eigenfunctions with quantum numbers $s=1/2$ and $m_s=\pm 1/2$ represented by 
two dimensional unit vectors and  
spherical harmonics with quantum numbers $\ell$ and $m_{\ell}$.
The pseudopotentials $v_{\ell j}^{PP}$ are constructed in such a way 
that the valence orbitals resulting from a 
pseudopotential  EXX calculation for the atom, i.e., the atomic pseudoorbitals
$\phi_{lj}$, are identical to the large components of the valence orbitals 
from a fully relativistic all electron
EXX calculation beyond the radius $r_c$ and have the same eigenvalues. 

The KS Hamiltonian operators of the pseudopotential EXX calculation 
for the atom and of the final plane wave EXX-SCDFT
calculation are nonrelativistic except 
that they contain the pseudopotentials $v_{\ell j}^{PP}$. The latter introduce
the relativistic effects in particular the spin-orbit interaction.
As a result the plane wave EXX-SCDFT calculation then differs
from a standard nonrelativistic pseudopotential EXX calculation not only 
by the fact that the orbitals are two-dimensional spinors that contain 
spin-up and spin-down components 
which, in general, both are nonvanishing, but in addition by the coupling
of the two spin components through the the $\ell j$-dependent pseudopotentials.
The latter are obtained 
along the lines described in Ref.\ [\onlinecite{engel01,engel01b}], however, 
without averaging the two $lj$-dependent pseudopotentials 
$v_{lj}^{PP}$ with $j=\ell-1/2$ and $j=\ell+1/2$.

Technically we construct $lj$-dependent pseudopotentials $v_{\ell j}^{PP}$
for a fully relativistic four-component pseudopotential calculation. 
The corresponding pseudoorbitals occuring in the separable
pseudopotentials $v_{lj}^{sep}$ of Eq. (\ref{pseudopotential}), at first, 
are four-component spinors. 
We then use only the renormalized large
components $\phi_{lj}$ of the original four-component pseudoorbitals 
to construct
separable pseudopotentials $v_{lj}^{sep}$ for a two-component 
valence-electron calculation that
takes into account relativistic effects only through the pseudopotentials. 
Strictly speaking this is somewhat inconsistent. However, 
the charge of the pseudoatoms is very low, for Si and Ge just 4 a.u. and
it is save to assume that direct relativistic effects, i.e., 
relativistic effects that do not have their origin
in the pseudopotential, but would result from a fully relativistic
four-component treatment of the valence electrons, 
are negligible for such small effective nuclear
charges. Indeed for the atoms Si and Ge the energetic differences 
from a two-component pseudopotential calculation and a fully
relativistic four-component pseudopotential calculation are of the order of 
$\mu$Hartree and thus completely negligible.

In order to analyze how relativistic effects are taken into account in a EXX
SCDFT calculation via the $j$-dependent pseudopotentials $v_{\ell j}^{PP}$
constructed with the relativistic EXX approach \cite{engel01,engel01b}, 
we consider for the moment the KS orbitals of a bulk system 
as linear combinations of atomic pseudoorbitals $\phi_{lj}$. One effect of
the relativistic pseudopotentials is that $s$- and $p$-orbitals are
contracted and lowered in energy while $d$- and $f$-orbitals are more diffuse
and raised in energy. The influence of these effects on the KS orbitals in
the bulk are automatically taken into account in a standard EXX treatment.
The $j$-dependence of the pseudopotentials, which is a consequence of
spin-orbit interactions, on the other hand, as mentioned above, 
leads to two-dimensional spinors as orbitals and causes features not
present in standard EXX methods, like spin-currents.  

For a more detailed discussion of the effect of spin-orbit interactions and
in particular of the generation of spin-currents we consider 
a two-component all-electron Pauli-Hamiltonian operator of an 
electronic system in a magnetic field including
spin-orbit interactions:
\begin{eqnarray}
\hat{H} \!\!\!\! 
&=& \hat{T} + \hat{V}_{ee} + \hat{v}_{ext} +  \hat{H}^{mag} 
+ \hat{H}^{SO}
\nonumber \\
&=&\hat{T} + \hat{V}_{ee} + \sum_{i=1}^N \bigg [ v_{ext}({\bf r}_i) 
+ {\textstyle \frac{1}{2}} {\bf p}_i \!\cdot\! {\bf A}({\bf r}_i)
+ {\textstyle \frac{1}{2}} {\bf A}({\bf r}_i) \!\cdot\! {\bf p}_i
\nonumber \\[-0.5em]
&&  \!\!\!\!\!\!\!\!
+ {\textstyle \frac{1}{2}} 
{\bf A}({\bf r}_i) \!\cdot\! {\bf A}({\bf r}_i)
+  {\textstyle\frac{1}{2}} \, \boldsymbol{\sigma} \!\cdot\! {\bf B}({\bf r}_i) 
+{\textstyle\frac{1}{4c^2}} \, \boldsymbol{\sigma} \!\cdot\! 
\left[ (\nabla v_{ext})({\bf r}_i)  \times {\bf p}_i \right] \!\!
\bigg ]
\nonumber \\
&=&\hat{T} + \hat{V}_{ee} + \int \! d{\bf r} \;\; \boldsymbol{\Sigma}^T \;
{\bf V}({\bf r}) \; \hat{\bf J}({\bf r}) \,.
\label{ham_all_el}
\end{eqnarray}
In Eq.\ [\ref{ham_all_el}], $\hat{T}$ and $\hat{V}_{ee}$ are the operators of
the kinetic energy and the electron-electron repulsion,  $\hat{v}_{ext}$ is
the operator generated by the external electrostatic potential ${v}_{ext}$, 
usually the potential of the nuclei, 
$\hat{H}^{mag}$ are the parts of the Hamiltonian operator caused 
by a magnetic field ${\bf B}$ 
with accompanying vector potential ${\bf A}$, and $\hat{H}^{SO}$ describes the
spin-orbit interaction. By ${\bf r}_i$ the position of
the $i$-th electron is denoted, by ${\bf p}_i$ the corresponding canonical
momentum operator. The vector $\boldsymbol{\sigma}$ contains the
Pauli spin matrices. The sum in the first line of Eq. [\ref{ham_all_el}] runs
over all $N$ electrons. The vector $\boldsymbol{\Sigma}$ has four components,
${\Sigma}_0$ being a 2x2 unit matrix and ${\Sigma}_1$, ${\Sigma}_2$, and
${\Sigma}_3$ being the Pauli spin matrices ${\boldsymbol\sigma}_x$, 
${\boldsymbol\sigma}_y$, and ${\boldsymbol\sigma}_z$. The four components 
of the vector $\hat{\bf J}({\bf r})$ are the density operator 
$\sum_{i=1}^N \; \delta({\bf r}-{\bf r}_i) = \hat{J}_{0}({\bf r})$
and the x-, y, z- components of the current operator   
$\left(\frac{1}{2}\right) 
\sum_{i=1}^N \;\; p_{x,i} \; \delta({\bf r}-{\bf r}_i) \;+\;
\delta({\bf r}-{\bf r}_i) \; p_{x,i} =  \hat{J}_{1}({\bf r})$ etc. 
The 4x4 matrix ${\bf V}({\bf r})$ is composed of matrixelements
${V}_{\mu\nu}({\bf r})$ with $\mu,\nu= 0,1,2,3$ and is given by
\begin{equation}
\mathbf{V}({\bf r})
=\!
\left( \!\!  \begin{array}{cccc}
v_{ext}({\bf r}) \!+\!  \frac{{\bf A}^2({\bf r})}{2} \!&\! 
A_x({\bf r}) \!&\! A_y({\bf r}) 
\!&\! A_z({\bf r}) \\
\frac{B_x({\bf r})}{2} \!\!\!&\!\!\! 0 
\!\!\!&\!\!\! -\frac{v_{ext,z}({\bf r})}{4c^2} 
&  \frac{v_{ext,y}({\bf r})}{4c^2} \\
\frac{B_y({\bf r})}{2} \!\!\!&\!\!\! \frac{v_{ext,z}({\bf r})}{4c^2} 
\!\!\!&\!\!\! 0 \!\!\!&\!\!\! -\frac{v_{ext,x}({\bf r})}{4c^2} \\
\frac{B_z({\bf r})}{2} \!\!\!&\!\!\! -\frac{v_{ext,y}({\bf r})}{4c^2}
&  \frac{v_{ext,x}({\bf r})}{4c^2} \!\!\!&\!\!\! 0
\end{array} \! \right) \,.
\label{V}
\end{equation}
In Eq.\ [\ref{V}], 
$v_{ext,x}({\bf r}) = \partial v_{ext}({\bf r}) / \partial x$. The 
fact that the spin-orbit interaction Hamiltonian operator $\hat{H}^{SO}$ 
can be expressed according to the last line of Eq.\ [\ref{ham_all_el}] together
with Eq.\ [\ref{V}] is crucial, because this represents the basis for the
SCDFT recently introduced in Ref.\ [\onlinecite{rohra06}]. 
The spin-orbit interaction only leads 
to additional terms in the matrix $\mathbf{V}$ of Eq.\ [\ref{V}], which do not
interfere with the derivation of the SCDFT of Ref.\ [\onlinecite{rohra06}],
i.e., the latter
derivation also holds in the presence of the spin-orbit Hamiltonian operator
$\hat{H}^{SO}$. 

According to Ref.\ [\onlinecite{rohra06}] the many-electron KS equation 
associated with a Schr\"odinger equation
with an Hamiltonian operator given by Eq.\ [\ref{ham_all_el}] 
is determined by the KS Hamiltonian operator
\begin{eqnarray}
\hat{H}_s \!\! &=& \hat{T} 
+ \int \! d{\bf r} \;\; \boldsymbol{\Sigma}^T \;
{\bf V}_s({\bf r}) \; \hat{\bf J}({\bf r})
\label{KS-Hamiltonian}
\end{eqnarray}
with the 4x4 matrix ${\bf V}_s$ representing the KS potential,
which is given by
${\bf V}_s({\bf r}) = {\bf V}({\bf r}) + {\bf U}({\bf r})
+ {\bf V}_{xc}({\bf r})$. The Hartree potential ${\bf U}$
and the exchange-correlation potential ${\bf V}_{xc}$ can be represented by
4x4 matrices. Within the matrix ${\bf U}$ only the component 
$U_{00}({\bf r}) = 
\int \!\! d{\bf r}' \frac{\rho_{00}({\bf r}')}{|{\bf r} -{\bf r}'|}$, 
the standard Hartree potential, is different from zero. 
The components 
${\bf V}_{xc,\mu\nu}$ of the 
exchange-correlation potential ${\bf V}_{xc}$ are given by
\begin{eqnarray}
V_{xc,\mu\nu}({\bf r}) = 
\frac{\delta E_{xc}}{\delta \rho_{\mu\nu}({\bf r})} \,.
\label{Vxc}
\end{eqnarray}

Within the considered SCDFT all 16 components of the spin-current density 
$\boldsymbol{\rho}$ of the KS system equal those of the real 
electronic system.
The components ${\rho}_{\mu\nu}$ with $\mu,\nu = 0,1,2,3$  
of $\boldsymbol{\rho}$ are defined as follows:
${\rho}_{00}$ is just the regular
ground state electron density, ${\rho}_{\mu 0}$ with $\mu = 1,2,3$ represent
the x-, y-, and z-components of the spin-density, ${\rho}_{0 \nu}$ with 
$\nu = 1,2,3$ represent the x-, y-, and z-components of the paramagnetic
current of the electron density, while  ${\rho}_{\mu\nu}$ 
with $\mu,\nu = 1,2,3$
represent the components of the paramagnetic currents of the
spin-density, i.e., currents of the magnetization. 

In the absence of
spin-orbit interactions a correct description of all 16 
components of the spin-current density $\boldsymbol{\rho}$ through the KS
system is desirable for
systems that exhibit non-vanishing currents of the spin-density, similarly
as it is desirable in standard, i.e., non-current, 
density-functional theory to treat 
spin-polarized systems within spin-density-functional theory in order to
correctly describe both spin-up and spin-down electron densities. However, 
in the absence of spin-orbit interactions, the Hamiltonian operator
Eq. (\ref{ham_all_el}) does not couple to currents of the spin-density, i.e.,
does not couple to ${\rho}_{\mu\nu}$ 
with $\mu,\nu = 1,2,3$. Therefore it is possible in these cases to apply the
standard current spin-density formalism of Vignale and Rasolt, 
which considers only the
ground state electron density, ${\rho}_{00}$, the spin-density
${\rho}_{\mu 0}$ with $\mu = 1,2,3$, and the paramagnetic
current of the electron density ${\rho}_{0 \nu}$ with 
$\nu = 1,2,3$, similarly as it is possible to treat spin-polarized systems 
via non-spin-polarized density-functional theory although this might affect
the accuracy in practical application, which require the use of
approximate density-functionals. In the presence of spin-orbit interactions the
situation is different. Because the Hamiltonian operator [\ref{ham_all_el}]
now couples to components of the paramagnetic currents of the
spin-density it is mandatory that the KS system 
correctly describes all 16 components
of the spin-current density ${\rho}_{\mu\nu}$, i.e., the standard current 
spin-density functional formalism is no longer applicable in the presence of
spin-orbit interactions.

In our case we do not treat spin-orbit interactions by the explicit spin-orbit
term $H^{SO}$ of Eq.\ (\ref{ham_all_el}),
but via the $j$-dependent pseudopotentials $v_{lj}^{sep}$ of
Eq. (\ref{pseudopotential}). 
However, $j$-dependent pseudopotentials, by construction, generate an
effect corresponding to that of the spin-orbit term $H^{SO}$ and thus
like the latter generate spin-currents and couple to them. It is important to
notice that spin-currents, due to the continuity of orbitals, are present 
not only close to the nuclei, but also in
the interstitial region between the atoms despite the fact that they 
are generated close to the nuclei by pseudopotentials. 
Thus spin-currents affect the bonding and
generally the electronic structure. A full and self-consistent treatment
of the effects of spin-currents in DFT therefore requires a spin-current
density-functional treatment as provided by the EXX SCDFT method of this work.

Note that an explicit treatment of spin-orbit interactions
via $H^{SO}$ in a selfconsistent KS method would be problematic,
because the spin-orbit term $H^{SO}$ arising from the 
relativistic many-electron
Hamiltonian operator via the singular Wouldy-Wouthausen transformation leads
to severe instabilities. Two-component methods that do not employ
pseudopotentials therefore treat spin-orbit effects either perturbatively 
\cite{ZORA} or within methods based on two-component Hamiltonian operators
obtained by other transformation like the
Douglas-Kroll-Hess transformation \cite{DKH}.

In the proposed combination of spin-orbit resolved $j$-dependent
pseudopotentials with the EXX-SCDFT method \cite{rohra06} we treat 
the exchange contribution ${\bf V}_{x}$
of the required exchange-correlation potential ${\bf V}_{xc}$ exactly and 
neglect the  correlation contribution. An inclusion of an approximate 
LDA or GGA correlation potential for the component ${V}_{c,00}$ would be 
straightforward. However, experience from standard EXX calculations shows 
that the inclusion of an LDA or GGA correlation potential has only 
marginal effect on the bandstructure. 

For the self consistent solid state calculations plane wave energy
cutoffs of 25 Ry for the orbitals and 14 Ry for the KS response function
were chosen. The set of used {\bf k}-points was chosen as
a uniform $4\times4\times4$ mesh covering the first Brillouin zone.
For the lattice constants of silicon and germanium 
the experimantal values of 5.4307\AA \, and 5.6579\AA \,, respectively,
were used.
Throughout our calculations, all conduction states were taken into account
for the construction of the response function and the right hand side of
the EXX equation \cite{stadele97,rohra06}. 
Pseudopotentials with angular momenta $l=0,1,2$ and 
with the $j$-quantum number taking the values $j\!=\!l\!+\!1/2$ and
$j\!=\!l\!-\!1/2$ were employed for both silicon and germanium.
Cutoff radii, in atomic units, 
of $r_{c,l=0}^{Si}\!=\!1.8$, $r_{c,l=1}^{Si}\!=\!2.0$ and
$r_{c,l=2}^{Si}\!=\!2.0$ for silicon, as well as $r_{c,l=0}^{Ge}\!=\!1.8$
$r_{c,l=1}^{Ge}\!=\!2.0$ and $r_{c,l=2}^{Ge}\!=\!2.9$ for germanium were used.
The  germanium $3d$-electrons were not included in the valence space.

\begin{figure}[t]
\includegraphics*[width=7.0cm]{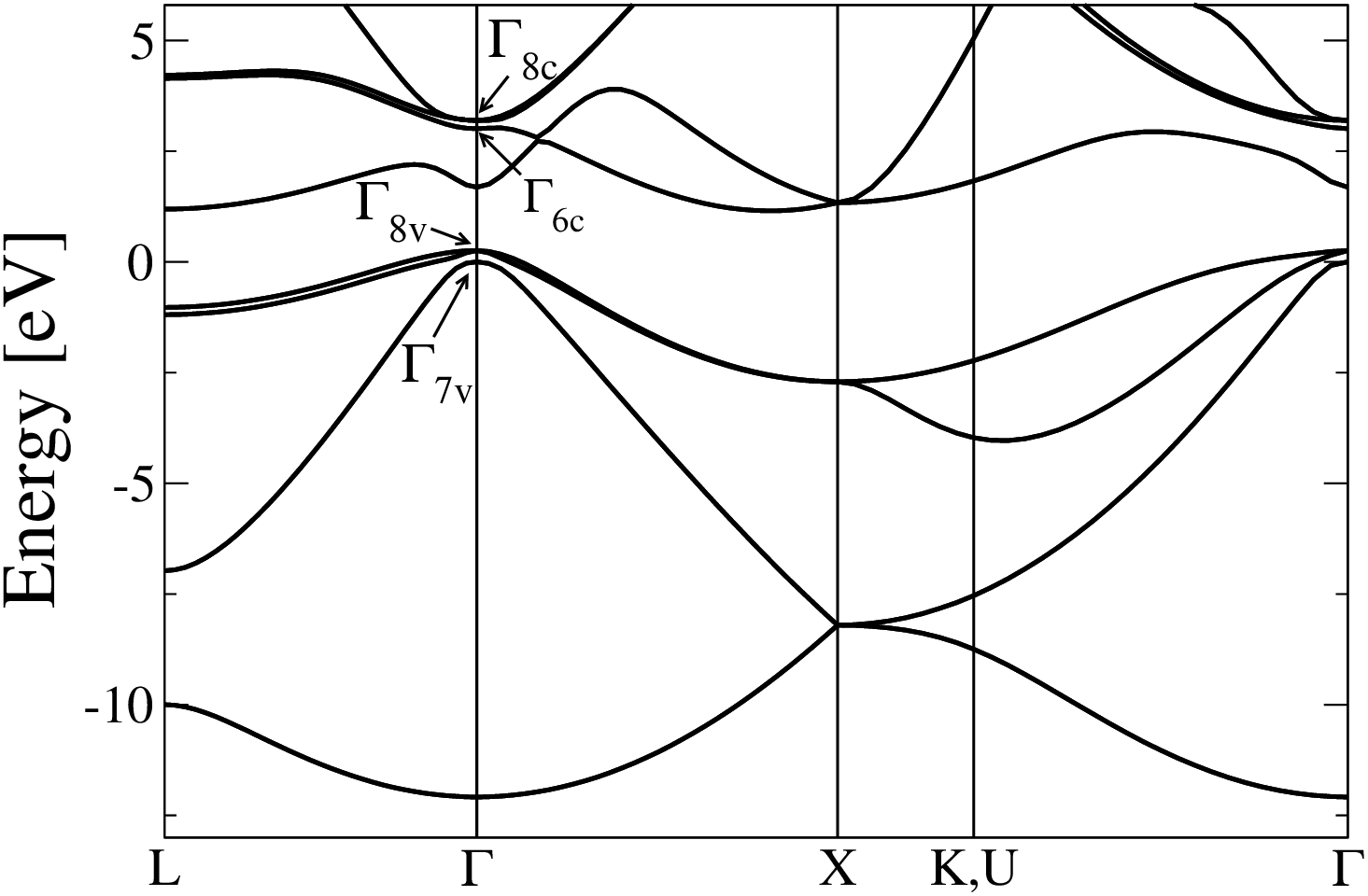}
\caption
{Bandstructure of germanium including spin-orbit coupling}
\label{fig.gebands}
\end{figure}

\begin{figure}[t]
\begin{center}
\subfigure
{\includegraphics[width=5.0cm]{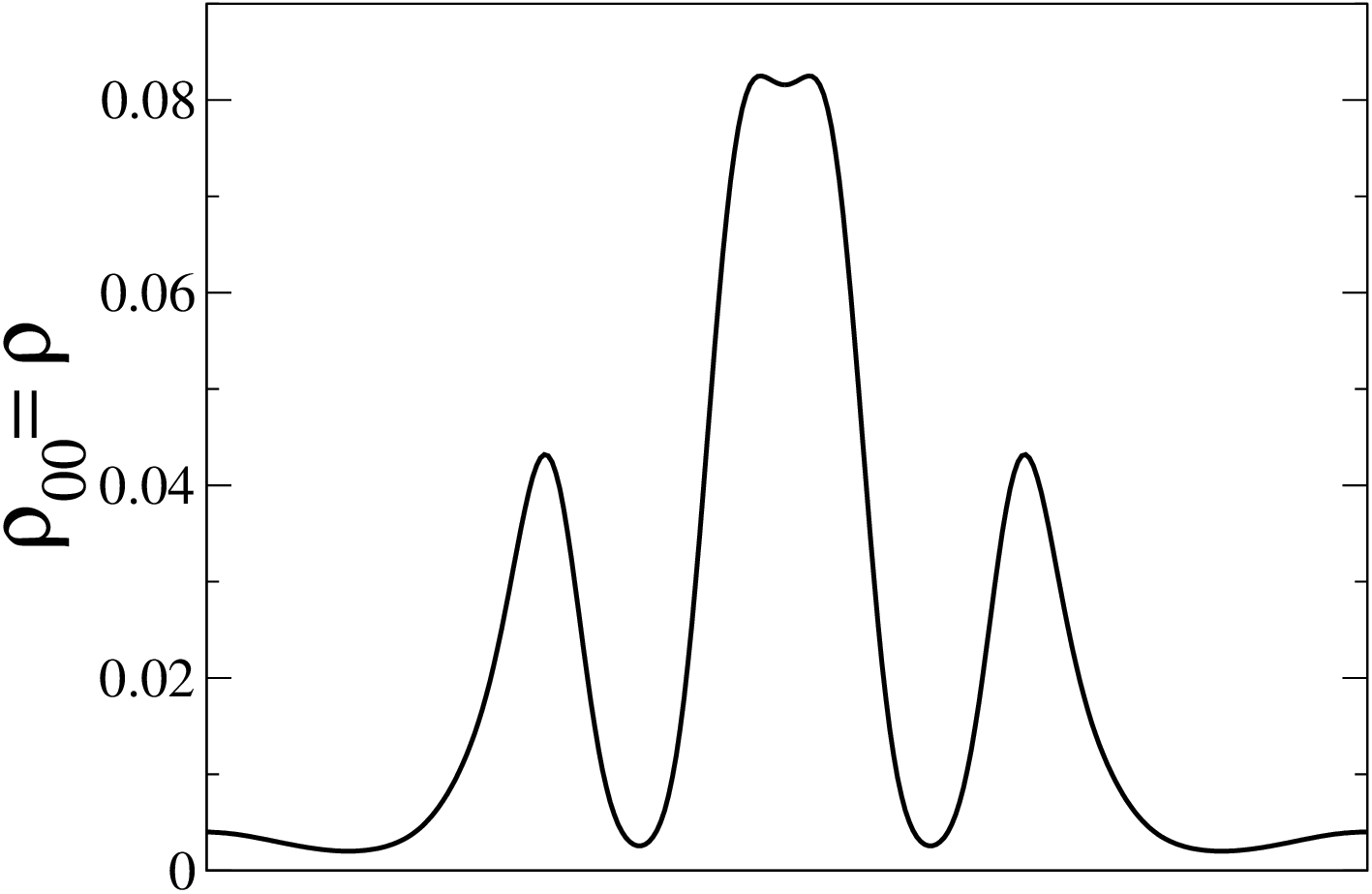}}
\end{center}
\vspace{-0.3cm}
\begin{center}
\subfigure
{\includegraphics[width=5.0cm]{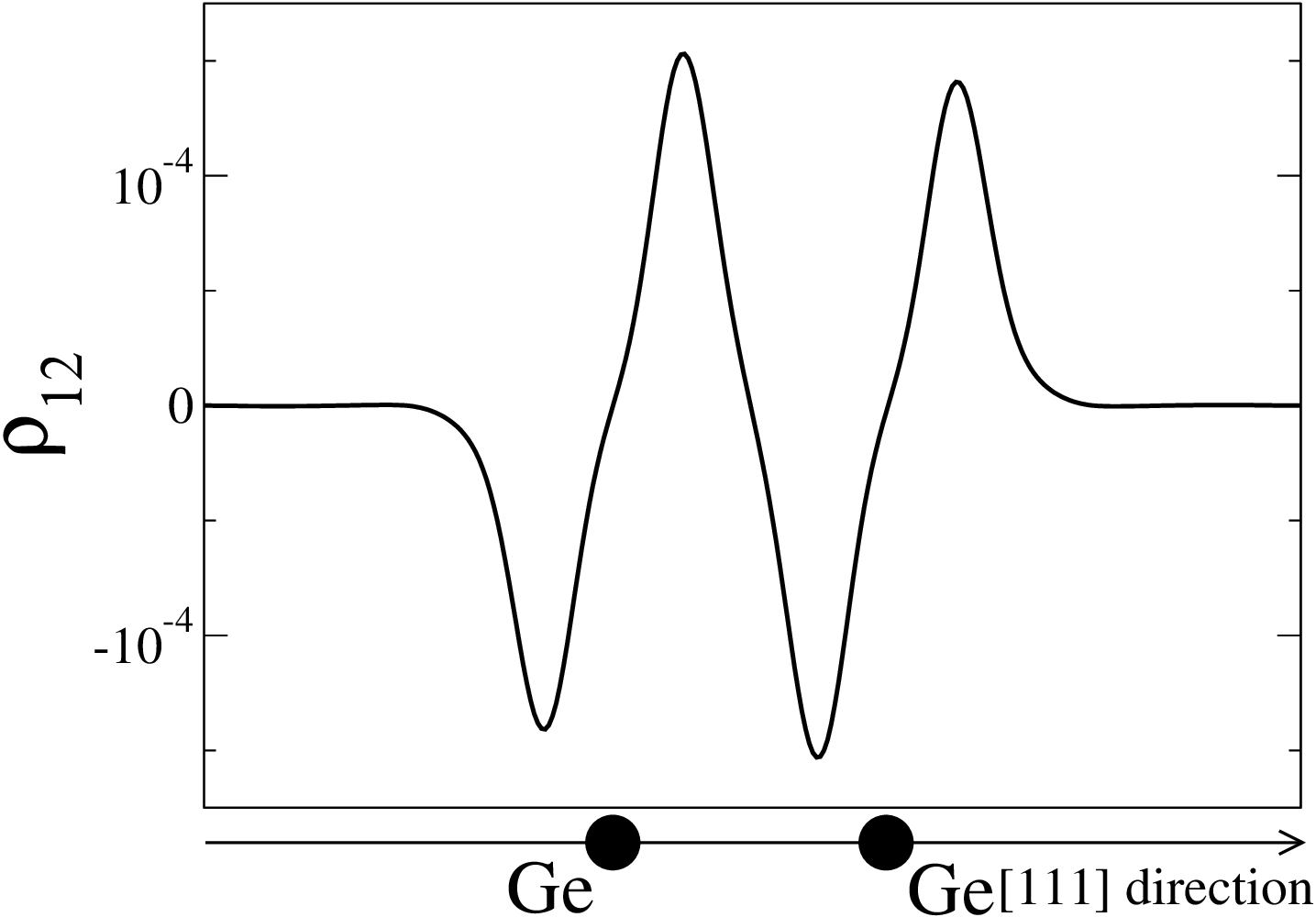}}
\end{center}
\vspace{-0.6cm}
\caption
{Electron density $\rho_{00}$ and spin-current $\rho_{12}$ of germanium 
along the bond axis ([111] direction)}
\label{fig.currents}
\end{figure}

The resulting EXX bandstructure for germanium is shown in
Fig.\ \ref{fig.gebands} with the values for the indicated
spin-orbit splittings listed in Table \ref{tab.sosplitge}.
The deviation of the calculated SCDFT spin-orbit splittings
from the experimental value is roughly 13.5\%. We believe that
this deviation can be attributed
to the fact that for the germanium calculations the 3$d$-states were
not included in the valence space. 
This explanation is supported by the results for silicon.
Here, no $d$-states are involved and the calculated spin-orbit splittings 
are in almost perfect agreement with experiment,
see Table \ref{tab.sosplitge}.
\begin{table}[h]
\caption{Spin-orbit splitting for germanium and silicon at the $\Gamma$-point
}
\begin{tabular}{p{3cm} p{3cm} c} 
\hline\hline
Germanium & \centering{$\Delta(\Gamma_{7v}-\Gamma_{8v})$} &
$\Delta(\Gamma_{6c}-\Gamma_{8c})$  \\ \hline
SCDFT                     & \centering{258.1} & 173.3 \\
Experiment\cite{madelung} & \centering{297} & 200 \\
\hline\hline
Silicon & \centering{$\Delta(\Gamma_{25v})$} &  \\ 
\hline
SCDFT                         & \centering{42.5} & \\
Experiment\cite{madelung}     & \centering{44.1} & \\
\hline\hline
\end{tabular}
\label{tab.sosplitge}
\end{table}


Fig.\ \ref{fig.currents} shows the electron density $\rho_{00}$ and
the spin-current $\rho_{12}$ along the bond axis, i.e., 
the unit cell's diagonal.
The plot for $\rho_{00}$ shows the usual minimum
of the valence electron density at the positions of the germanium nuclei. The
currents $\rho_{0\mu}$, the spin-densities $\rho_{\mu 0}$, 
and the spin-currents $\rho_{\mu\mu}$
(for $\mu = 1,2,3$) turned out to be zero. However, the displayed spin-current
$\rho_{12}$ differs from zero and the spin-currents 
$\rho_{23}$ and $\rho_{31}$ equal $\rho_{12}$ and thus also are non-zero. 
{\em This shows that spin-orbit effects indeed
induce spin-currents.}  The latter influence
the resulting bandstructures and therefore should not be neglected in
a selfconsistent treatment of spin-orbit effects.
Furthermore, the symmetry of the investigated system proposes the relation
$\rho_{12} = \rho_{23}= -\rho_{13} = -\rho_{21} = -\rho_{32}= \rho_{31}$
between the spin-currents with nonzero value. This relation was confirmed
by the results of our calculations.

In summary, we have presented a method to treat non-collinear spin, magnetic 
effects, currents, spin-currents, and spin-orbit interactions in solids 
on an equal footing in a self-consistent EXX Kohn-Sham approach and 
showed that spin-orbit coupling induces spin-currents.



\end{document}